# Mapping the unconventional orbital texture in topological crystalline insulators

Ilija Zeljkovic[*,1], Yoshinori Okada[*,1,2], Cheng-Yi Huang[3], R. Sankar[4], Daniel Walkup[1], Wenwen Zhou[1], Maksym Serbyn[5], Fangcheng Chou[4], Wei-Feng Tsai[3], Hsin Lin[6], A. Bansil[7], Liang Fu[5], M. Zahid Hasan[8] and Vidya Madhavan[1]

[1]Department of Physics, Boston College, Chestnut Hill, Massachusetts 02467, USA [2]WPI-AIMR, Tohoku University, Sendai, 980-8577, Japan [3]Department of Physics, National Sun Yat-sen University, Kaohsiung 80424, Taiwan [4]Center for Condensed Matter Sciences, National Taiwan University, Taipei 10617, Taiwan [5]Department of Physics, Massachusetts Institute of Technology, Cambridge, MA 02139, USA [6]Graphene Research Centre and Department of Physics, National University of Singapore, Singapore 117542 [7]Department of Physics, Northeastern University, Boston, Massachusetts 02115, USA [8]Joseph Henry Laboratory, Department of Physics, Princeton University, Princeton, New Jersey 08544, USA *These authors contributed equally to this work

**The newly discovered topological crystalline insulators (TCIs) harbor a complex band structure involving multiple Dirac cones[1–6]. These materials are potentially highly tunable by external electric field, temperature or strain and could find future applications in field-effect transistors, photodetectors, and nano-mechanical systems. Theoretically, it has been predicted that different Dirac cones, offset in energy and momentum-space, might harbor vastly different orbital character[7], a unique property which if experimentally realized, would present an ideal platform for accomplishing new spintronic devices. However, the orbital texture of the Dirac cones, which is of immense importance in determining a variety of materials' properties[8–13], still remains elusive in TCIs. Here, we unveil the orbital texture in a prototypical TCI $Pb_{1-x}Sn_xSe$. By using Fourier-transform (FT) scanning tunneling spectroscopy (STS) we measure the interference patterns produced by the scattering of surface state electrons. We discover that the intensity and energy dependences of FTs show distinct characteristics, which can directly be attributed to orbital effects. Our experiments reveal the complex band topology involving two Lifshitz transitions[14] and establish the orbital nature of the Dirac bands in this new class of topological materials, which could provide a different pathway towards future quantum applications.**

A counterpart to charge and spin, electron orbitals are of immense importance in the underlying physical processes of a variety of systems. The orbital degree of freedom, for example, plays a crucial role in the colossal magnetoresistance effect in manganese oxides, and contributes to the anisotropic electronic and magnetic properties in many other transition-metal oxide systems [8–10]. More recently, orbital ordering within the superconducting FeAs layer has been thought to govern structural phase transitions and "stripe"-like antiferromagnetism in Fe-based high-temperature superconductors [11–13]. Similarly, topological materials host complex orbital arrangements often strongly coupled to other electronic degrees of freedom [15–21]. TCIs in particular are predicted to exhibit intricate band, spin and orbital textures, potentially relevant for interactions in the quantum Hall regime. Although previous experiments provided a glimpse into the complex band topology present in TCIs [3–5,22,23], these experimental efforts have not been able to shed light onto its orbital-texture. Here we use Fourier-transform (FT) scanning tunneling spectroscopy (STS) to reveal the distinct orbital nature of the Dirac bands in the TCI, $Pb_{1-x}Sn_xSe$.

In its stoichiometric state, $Pb_{1-x}Sn_xSe$ with x=0 is a trivial insulator under the $Z_2$ topological classification of materials due to the absence of band inversion. The process of adding Sn, which substitutes for Pb, leads to band inversion at an even number of time-reversal points, and the solutions continue to remain $Z_2$ trivial. The emergence of topologically protected surface states however occurs at x>~0.23 due to the nontrivial band topology classified by crystalline symmetries.[3] $Pb_{1-x}Sn_xSe$ crystals cleave along (001) crystal direction based on the square lattice observed in STM topographs (Fig. 1a,b) and 3Å step heights (*Supplementary Discussion I*). We therefore consider the band structure at the (001) face. From theory, the SS band structure consists of two "parent" Dirac cones centered at X and vertically offset in energy[2,24]. When they intersect, the hybridization between the electron-branch of the lower parent Dirac cone and the hole-branch of the upper parent Dirac cone opens a gap at all points except along the mirror line, leading to the formation of a pair of lower-energy "child" Dirac points shifted away in momentum space from the high-symmetry point X[2,24] (Fig. 1c). In addition, the SS band structure exhibits two Lifshitz transitions[14] at energies $E_{VHS+}$ and $E_{VHS-}$ where the constant energy contours (CECs) in momentum space (*k*-space) change from two disjointed hole pockets located symmetrically with respect to X, to one large hole pocket and a smaller electron pocket, both centered at X (Figs. 1c,d).[2,24]

Figure 1e shows position dependent d*I*/d*V* spectra on this material. In general, the d*I*/d*V* spectra are V-shaped, with two well-defined peaks (Fig. 1e). From our previous studies, the minimum in the density of states of the V-shaped feature denotes the energy of the Dirac point and the two peaks on either side represent Van-Hove singularities ($E_{VHS+}$/ $E_{VHS-}$) associated with the Lifshitz transition.[25] Interestingly, despite the disordered nature of our samples with greater than 30% of the Pb sites replaced by Sn, we find that the Dirac point and the overall spectra remain homogeneous to within a few meV (Fig. 1e, *also see Supplementary Discussion II*). Having identified the important energy scales, we proceed to the application of the Fourier-transform quasiparticle interference (FT-QPI) on this system.

FT-QPI imaging has been successfully applied to extract the band structure of many complex systems, such as high-temperature superconductors,[26–28] heavy-fermion compounds[29,30] and $Z_2$ topological materials[31–33], but this technique is yet to be fully utilized for spin- and orbital-texture mapping of extracted bands. Representative FTs of *dI/dV* conductance maps acquired over a 1300Å square region of $Pb_{0.70}Sn_{0.30}Se$ are shown in Fig. 2. As a starting point to analyze the FTs the expected QPI pattern can be obtained using a simple autocorrelation of the CECs, not including any matrix element effects (Fig. 2b,e). The pattern produced can be visualized by studying the Fermi surface topology shown in Fig. 2a,d. From this, we see that two main sets of scattering wavevectors ($Q_1$/$Q_3$ and $Q_2$/$Q_4$) are expected for the (001) SS. Above the Lifshitz transition for example (Fig. 2d,e), the QPI pattern around the momentum positions $Q_1$/$Q_3$ originates from scattering between two (110) mirror-symmetric pairs of Dirac cones, and is expected to be quasi-elliptical. In contrast, the pattern at $Q_2$/$Q_4$ marks the scattering between two pairs of child Dirac cones rotated 90 degrees with respect to one other, and should be nearly circular (also see *Supplementary Discussion III*).

Our first observation is that there is a marked difference between the QPI patterns above $E_{VHS+}$ (Fig. 3a), where clear, nearly circular dispersing features appear, at $E_{VHS+}$ (Fig. 3b), and below $E_{VHS+}$, where a (non-dispersing) "clover"-like set of four dots appear (Fig. 3c). This remarkable switch in the QPI pattern is directly connected to the change in Fermi surface topology across the Lifshitz transition (Fig. 3a-f). The simple autocorrelation of the CECs shown in Fig. 2b,e confirms this picture; the disconnected CECs below $E_{VHS+}$ result in a non-dispersing clover pattern near $Q_2$/$Q_4$, while the continuous oval-shaped CECs above $E_{VHS+}$ result in a quasi-circular dispersing pattern.

The main challenge of any QPI study is extracting the band structure in *k*-space from the observed dispersive modes in *q*-space. An important piece of information is provided in the set of scattering wavevectors $Q_2$, close to the Dirac point where the clover is observed (Fig. 2a-c and Fig. 3c). Since the clover is a product of quasiparticle scattering between four child Dirac cones in the first BZ, the average distance in *q*-space between the centers of neighboring clover leafs allows us to precisely extract the *k*-space distance between two Dirac cones symmetric around X to be 0.060 ± 0.006 1/Å. Interestingly, as the *k*-space location of the Dirac cones in TCIs is directly connected to doping,[22] and nominal doping is often quite different from the actual and can vary within the sample, this method presents an alternative way to locally determine and compare the doping level in this class of materials.

In order to extract SS dispersion anisotropy in *k*-space, we need to use the *q*-space dispersions of the sets of wavevectors at *both* $Q_1$ and $Q_2$. Remarkably, although $Q_1$ is expected to be weak, the high quality of our data allows us to track $Q_1$ over a small range of energies (Fig. S4). Using the method explained in *Supplementary Discussion III,* we obtain the dispersions along Γ-X and X-M directions (Fig. 3g) and determine the velocities to be $v_x$=3.49eVÅ and $v_y$=2.20eVÅ, giving the anisotropy coefficient of 1.59 and the arithmetic mean of velocities to be $(v_x v_y)^{1/2}$= 2.77 eVÅ. It is important to note that no theoretical models were used to obtain these SS band dispersion velocities. Additionally, although these velocities have been calculated without any assumptions about the exact shape of the CECs *or* the positions of the Lifshitz transitions, the position of the upper Lifshitz transition from Fig. 3g matches beautifully with the energy of the corresponding feature in the average d*I*/d*V* spectrum obtained on the same sample (Fig. 3h). Furthermore, the arithmetic mean of dispersion velocities $(v_x v_y)^{1/2}$ from Landau levels in $Pb_{1-x}Sn_xSe$ is found to be 2.60 eVÅ and 2.70 eVÅ for x~0.33 and x~0.30 respectively[25], and is in excellent match with the value of 2.77 eVÅ obtained from QPI dispersion in x~0.37 sample.

Having determined the SS dispersion, we turn to one of the most striking aspects of the data, i.e., the asymmetry in the intensity of the QPI pattern across $E_d$, which can be observed by comparing Fig. 4a and Fig. 4c (also see *Supplementary Discussion IV*). As we show in the following discussion, this asymmetry is a direct consequence of the highly non-trivial orbital texture of the Dirac SS bands. The spin texture in this material can be simply stated as exhibiting left-handed chirality for the electron branches of all parent and child Dirac cones (Fig. 1d). The orbital character is however more complex. While the two child Dirac cones are necessarily related by time-reversal symmetry and mirror symmetry, the two parent Dirac cones do not have to be similar (each of the parent Dirac cone maps to itself under either the time reversal or mirror operator). In fact, by the symmetry arguments with the aid of first-principles band structure computations[7], the two parent Dirac cones are expected to have different orbital character associated with orbitals with opposite sign of mirror eigenvalues. Let us now consider the two sets of parent/child Dirac cones in *k*-space, one around X($\pi/a_0$,0) and the other one around Y(0, $\pi/a_0$) (Fig. 1c). Theoretically, while the upper Sn parent Dirac cone is associated with $p_z$ orbital character for both X- and Y-momenta, the lower Se parent Dirac cone has different orbital wavefunction, i.e. Se $p_x$ around X, and Se $p_y$ around Y. The different orbital textures below $E_d$ at ($\pi/a_0$,0) and (0, $\pi/a_0$) positions (Fig. 1c) should suppress scattering between them and therefore result in diminished QPI patterns below $E_d$, exactly as seen in Figure 4. Our data thus provide direct experimental confirmation of the proposed orbital arrangement of the TCI SS.

To further substantiate this picture, we use the proposed spin and orbital textures in model simulations of the QPI data. We find that the resulting simulations capture many features of our data, including changes in intensity with angle in the set of scattering wavevectors $Q_2$ (*Supplementary Discussion VI*), that were lacking in the simple autocorrelations. Take the QPI pattern at $E_d$+122meV as an example (Fig. 4d-f). The outmost high-intensity ring is due to the scattering between the two Sn $p_z$ parent cone across X- and Y-neighborhood. The inner ring comes from the scattering between Se parent cone at one BZ corner neighborhood and Sn cone at the other. Without the spin- and orbital-matrix element effect (Fig. 4e), little intensity variation around the rings was found in the autocorrelation map. When the matrix element effect is turned on (Fig. 4f), the inner ring is much suppressed since it represents scattering between different kinds of orbitals, once again consistent with the postulated orbital texture. We note here that the intensity variation on the outer ring is in part due to the spin-texture on the oval-shaped CEC and partly a consequence of the chosen impurity potential for matching the experimental data (*Supplementary Discussion VI*).

Finally, our data show evidence for asymmetry in dispersion velocities between hole- and electron-branches of the outer Dirac cones. The dashed gray lines in Fig. 2g represent linear fits to the experimental data points and show a slight asymmetry in the slope above and below $E_d$. This asymmetry is also noticeable in the Landau level dispersion in the same material.[25] Even though it is plausible that such asymmetry could occur due to proximity of SS to bulk bands which might be affecting the dispersion differently in the two energy regimes, the upper and lower Lifshitz transitions in the same samples are also asymmetric with respect to the Dirac point (*Supplementary Discussion V)*. Our observations provide strong evidence that, in addition to asymmetric nature of the orbital wavefunction, a particle-hole symmetric model with identical parent Dirac cones dispersions may not be enough to completely encompass the underlying physics present in this class of materials and that non-identical parent Dirac cones need to be taken into account.

## Methods

$Pb_{1-x}Sn_xSe$ single crystals used for QPI imaging were grown by self-selecting vapor growth method, cleaved at 77 K, and immediately inserted into the STM head. Doping concentration was checked using energy-dispersive X-ray spectroscopy (EDS). All d*I*/d*V* measurements were acquired at 6 K using a standard lock-in technique with 1488 Hz frequency. We use Lawler-Fujita drift-correction algorithm[34] on all acquired data to remove the effects of slow thermal and piezoelectric drift. Quasiparticle interference (QPI) imaging technique we used utilizes the interference of elastically scattered quasiparticles (which are just electrons in this case) with different momenta $k_1$ and $k_2$, resulting in a standing wave "ripples" of wavevector $q=k_1-k_2$ which can be detected in STM d*I*/d*V* conductance maps. Even though QPI patterns can be easily seen in real-space in a prototypical metal such as Cu(111),[35] two-dimensional FTs of d*I*/d*V* maps (FT-QPI) have proven to be necessary to extract all the scattering wave vectors in more complicated systems.[26]

## References

1. Fu, L. Topological Crystalline Insulators. *Physical Review Letters* **106**, 106802 (2011).
2. Hsieh, T. H. *et al.* Topological crystalline insulators in the SnTe material class. *Nature Communications* **3**, 982 (2012).
3. Dziawa, P. *et al.* Topological crystalline insulator states in $Pb_{1-x}Sn_xSe$. *Nature Materials* **11**, 1023–7 (2012).
4. Xu, S.-Y. *et al.* Observation of a topological crystalline insulator phase and topological phase transition in $Pb_{1-x}Sn_xTe$. *Nature Communications* **3**, 1192 (2012).
5. Tanaka, Y. *et al.* Experimental realization of a topological crystalline insulator in SnTe. *Nature Physics* **8**, 800–803 (2012).
6. Slager, R.-J., Mesaros, A., Juričić, V. & Zaanen, J. The space group classification of topological band-insulators. *Nature Physics* **9**, 98–102 (2012).

7. Wang, Y. J. *et al.* Nontrivial spin texture of the coaxial Dirac cones on the surface of topological crystalline insulator SnTe. *Physical Review B* **87**, 235317 (2013).

8. Tokura, Y. & Nagaosa, N. Orbital Physics in Transition-Metal Oxides. *Science* **288**, 462–468 (2000).

9. Chuang, Y. D., Gromko, a D., Dessau, D. S., Kimura, T. & Tokura, Y. Fermi surface nesting and nanoscale fluctuating charge/orbital ordering in colossal magnetoresistive oxides. *Science* **292**, 1509–13 (2001).

10. Khalifah, P. *et al.* Orbital ordering transition in $La_4Ru_2O_{10}$. *Science* **297**, 2237–40 (2002).

11. Krüger, F., Kumar, S., Zaanen, J. & Van den Brink, J. Spin-orbital frustrations and anomalous metallic state in iron-pnictide superconductors. *Physical Review B* **79**, 054504 (2009).

12. Lv, W., Wu, J. & Phillips, P. Orbital ordering induces structural phase transition and the resistivity anomaly in iron pnictides. *Physical Review B* **80**, 224506 (2009).

13. Lee, C.-C., Yin, W.-G. & Ku, W. Ferro-Orbital Order and Strong Magnetic Anisotropy in the Parent Compounds of Iron-Pnictide Superconductors. *Physical Review Letters* **103**, (2009).

14. Lifshitz, I. Anomalies of electron characteristics of a metal in the high pressure region. *Sov. Phys. JETP* **11**, 1130–1135 (1960).

15. Fu, L., Kane, C. & Mele, E. Topological Insulators in Three Dimensions. *Physical Review Letters* **98**, 106803 (2007).

16. Hasan, M. Z. & Kane, C. L. Colloquium: Topological insulators. *Reviews of Modern Physics* **82**, 3045–3067 (2010).

17. Qi, X.-L. & Zhang, S.-C. Topological insulators and superconductors. *Reviews of Modern Physics* **83**, 1057–1110 (2011).

18. Hsieh, D. *et al.* A tunable topological insulator in the spin helical Dirac transport regime. *Nature* **460**, 1101–5 (2009).

19. Cao, Y. *et al.* Mapping the orbital wavefunction of the surface states in three-dimensional topological insulators. *Nature Physics* **9**, 499–504 (2013).

20. Chen, Y. L. *et al.* Experimental realization of a three-dimensional topological insulator, $Bi_2Te_3$. *Science* **325**, 178–81 (2009).

21. Xia, Y. *et al.* Observation of a large-gap topological-insulator class with a single Dirac cone on the surface. *Nature Physics* **5**, 398–402 (2009).

22. Tanaka, Y. *et al.* Tunability of the k-space location of the Dirac cones in the topological crystalline insulator $Pb_{1-x}Sn_xTe$. *Physical Review B* **87**, 155105 (2013).

23. Gyenis, A. *et al.* Quasiparticle interference on the surface of the topological crystalline insulator $Pb_{1-x}Sn_xSe$. *Physical Review B* **88**, 125414 (2013).

24. Liu, J., Duan, W. & Fu, L. Surface States of Topological Crystalline Insulators in IV-VI Semiconductors. *Arxiv Preprint* 1304.0430 (2013).

25. Okada, Y. *et al.* Observation of Dirac node formation and mass acquisition in a topological crystalline insulator. *Science* **341**, 1496–9 (2013).

26. Hoffman, J. E. *et al.* Imaging quasiparticle interference in $Bi_2Sr_2CaCu_2O_{8+\delta}$. *Science* **297**, 1148–51 (2002).

27. Wang, Q.-H. & Lee, D.-H. Quasiparticle scattering interference in high-temperature superconductors. *Physical Review B* **67**, 020511 (2003).

28. Allan, M. P. *et al.* Anisotropic Energy Gaps of Iron-Based Superconductivity from Intraband Quasiparticle Interference in LiFeAs. *Science* **336**, 563–567 (2012).

29. Zhou, B. B. *et al.* Visualizing nodal heavy fermion superconductivity in $CeCoIn_5$. *Nature Physics* **9**, 474–479 (2013).

30. Allan, M. P. *et al.* Imaging Cooper pairing of heavy fermions in $CeCoIn_5$. *Nature Physics* **9**, 468–473 (2013).

31. Seo, J. *et al.* Transmission of topological surface states through surface barriers. *Nature* **466**, 343–6 (2010).

32. Okada, Y. *et al.* Direct Observation of Broken Time-Reversal Symmetry on the Surface of a Magnetically Doped Topological Insulator. *Physical Review Letters* **106**, 206805 (2011).

33. Okada, Y. *et al.* Ripple-modulated electronic structure of a 3D topological insulator. *Nature Communications* **3**, 1158 (2012).

34. Lawler, M. J. *et al.* Intra-unit-cell electronic nematicity of the high-$T_c$ copper-oxide pseudogap states. *Nature* **466**, 347–351 (2010).

35. Crommie, M. F., Lutz, C. P. & Eigler, D. M. Imaging standing waves in a two-dimensional electron gas. *Nature* **363**, 524–527 (1993).

# Figures

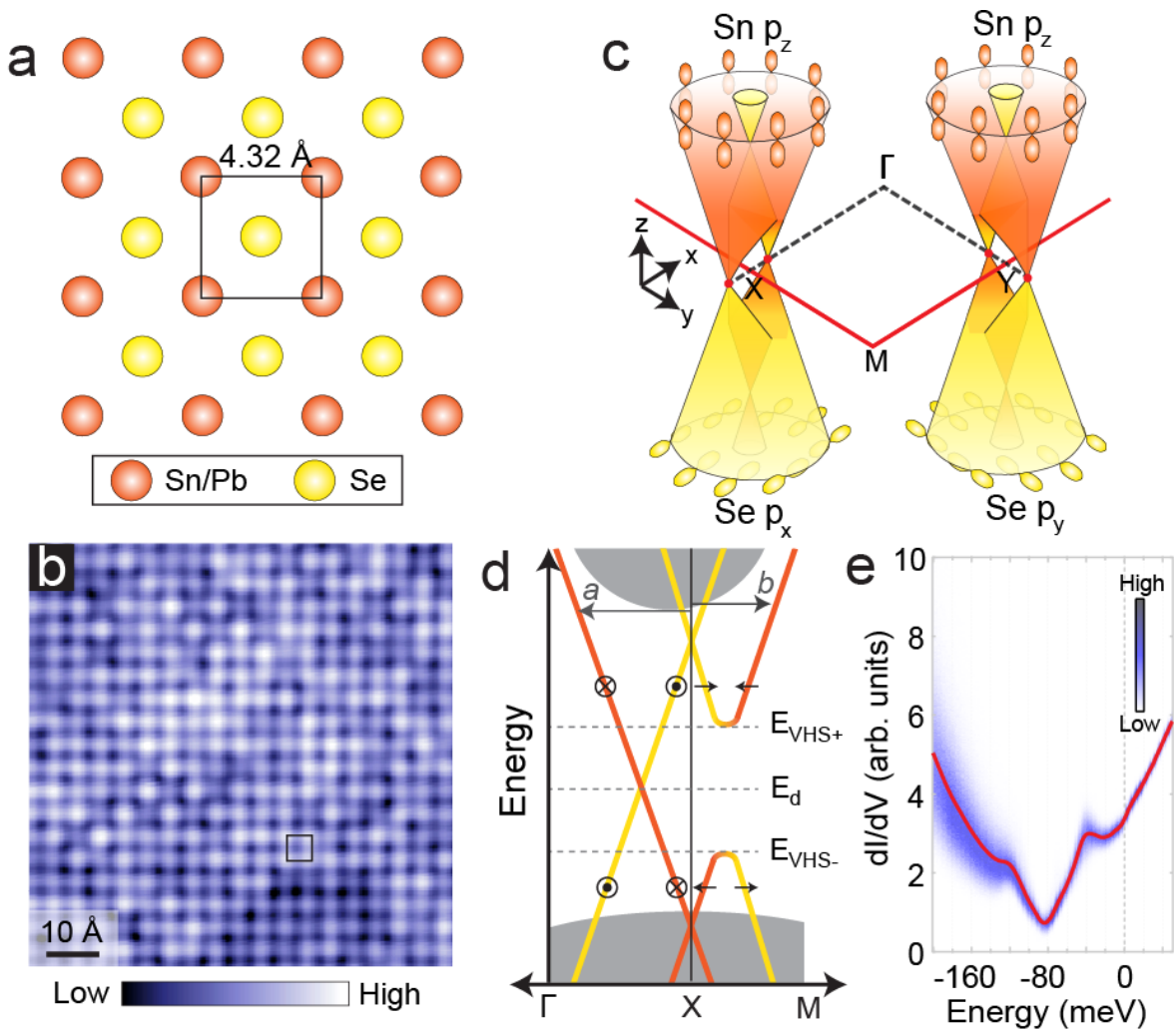

**Figure 1.** Overall band- and crystal-structure of topological crystalline insulator, $Pb_{1-x}Sn_xSe$. (a) Schematic representation of the (001) surface layer consisting of two inequivalent sublattices. (b) Typical STM topograph of $Pb_{1-x}Sn_xSe$ single crystals showing a square atomic lattice with ~4.32 Å periodicity ($I_{set}$=10 pA, $V_{set}$=-100 mV). The black square in (a,b) denotes the single sublattice seen in STM topographs. Schematic of the (c) orbital-texture of the two pairs of non-identical Dirac cones, and (d) band structure diagram of the surface states along Γ-X and X-M directions within a single pair of Dirac cones. Orange and yellow curves in (c,d) signal Sn and Se orbital character respectively, emphasizing different orbital texture across the Dirac point. Gray color in (d) represents bulk bands. Different spin directions are denoted along each branch in (d). (e) Typical average d$I$/d$V$ spectrum (red) and the distribution of d$I$/d$V$ spectra (blue) acquired across a ~30 nm square region, described in more detail in *Supplementary Discussion II* ($I_{set}$=135 pA, $V_{set}$=59 mV).

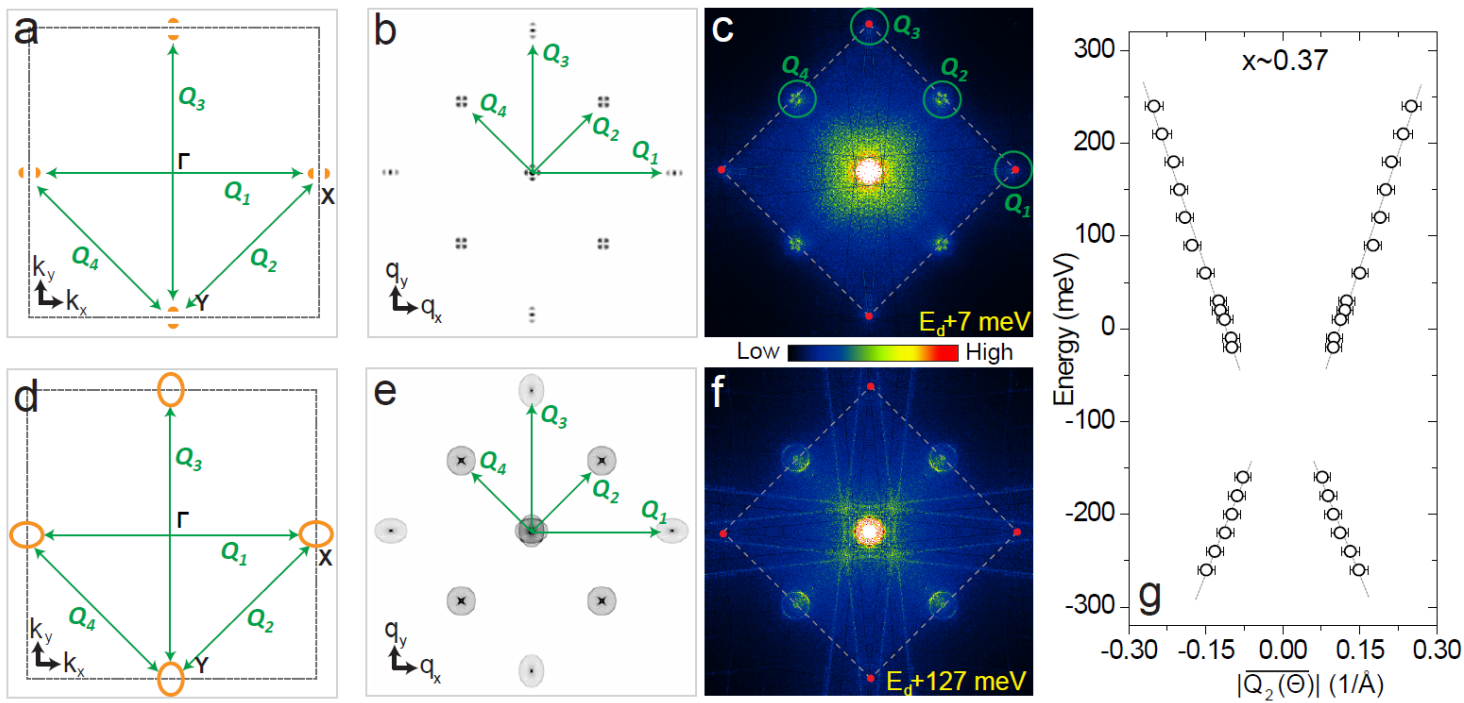

**Figure 2.** Representative *k*- and *q*-space structure and QPI dispersion. Schematic Fermi surface of TCI Pb$_{1-x}$Sn$_x$Se in the first tetragonal Brillouin zone (dashed square) at: (a) energy close to close to E$_d$, and (d) at arbitrary energy higher than E$_{VHS+}$. Orange lines represent the SS bands originating from four pairs of child Dirac cones centered at each X point. Quasiparticle scattering occurs between different pockets, and is expected to result in four principal scattering wavevectors (green arrows). (b,e) Autocorrelation of CECs in (a) and (d) respectively, showing the expected QPI signature. FTs of d*I*/d*V* conductance maps acquired at representative energies: (c) close to E$_d$, (f) above E$_{VHS+}$. (g) Energy dispersion of $\overline{|Q_2(\Theta)|}$ in Pb$_{0.63}$Sn$_{0.37}$Se. The dispersion is calculated from the set of scattering wave vectors Q$_2$ as: $\overline{|Q_2(\Theta)|} = \frac{1}{2\pi}\int_0^{2\pi}|Q_2(\Theta)|d\Theta$, where $Q_2(\Theta)$ is the angle-dependent radius of the quasi-circular feature, measured from $(\frac{\pi}{a_0}, \frac{\pi}{a_0})$ point in *q*-space as outlined in Fig. 3(a) (*Supplementary Discussion III*). Dashed lines in (g) represent linear fits to the data.

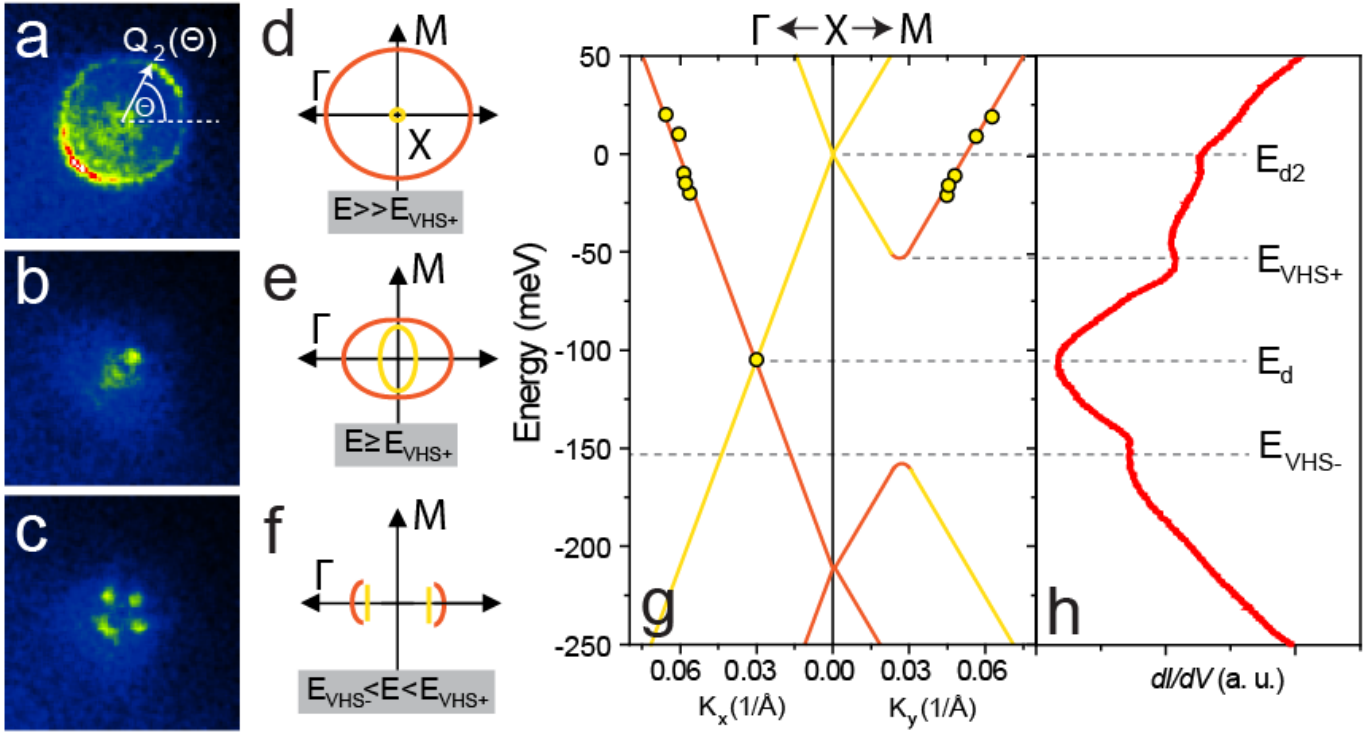

**Figure 3.** Visualizing the Lifshitz transition and extracting the anisotropy of the SS dispersion. Zoom-in on the set of scattering wave vectors $Q_2$ at energies: (a) $E_d$+122 meV, above $E_{VHS+}$ (b) $E_d$+52 meV, at $E_{VHS+}$ and (c) $E_d$+2 meV, below $E_{VHS+}$, emphasizing the striking change in the QPI structure across the Lifshitz transition. The schematic CEC for each of these energies is shown in (d), (e), and (f) respectively. (g) SS band structure along X-Γ and X-M directions of $Pb_{0.63}Sn_{0.37}Se$. Calculated dispersion velocities along X-Γ and X-M directions are *3.49 eVÅ* and *2.20 eVÅ* respectively, providing the first insight into the high dispersion anisotropy in these samples which is determined to be 1.59. Each data point (yellow circle) was extracted solely from the dispersion of the sets of scattering wavevectors $Q_1$ and $Q_2$ (*Supplementary Discussion III*). Solid orange and yellow lines in (g) are fits to the experimental data, extrapolated for visualization purposes assuming particle-hole symmetry. (h) Average *dI/dV* spectrum on the same sample over the same energy range as the plot in (g). Horizontal gray dashed lines across (g,h) denote the positions of prominent features in the spectrum ($E_d$, $E_{d2}$, $E_{VHS+}$, $E_{VHS-}$) and show an excellent match with the SS dispersion extracted from QPI. Orange and yellow lines in all panels represent Pb/Sn-like and Se-like orbital character of the states respectively.

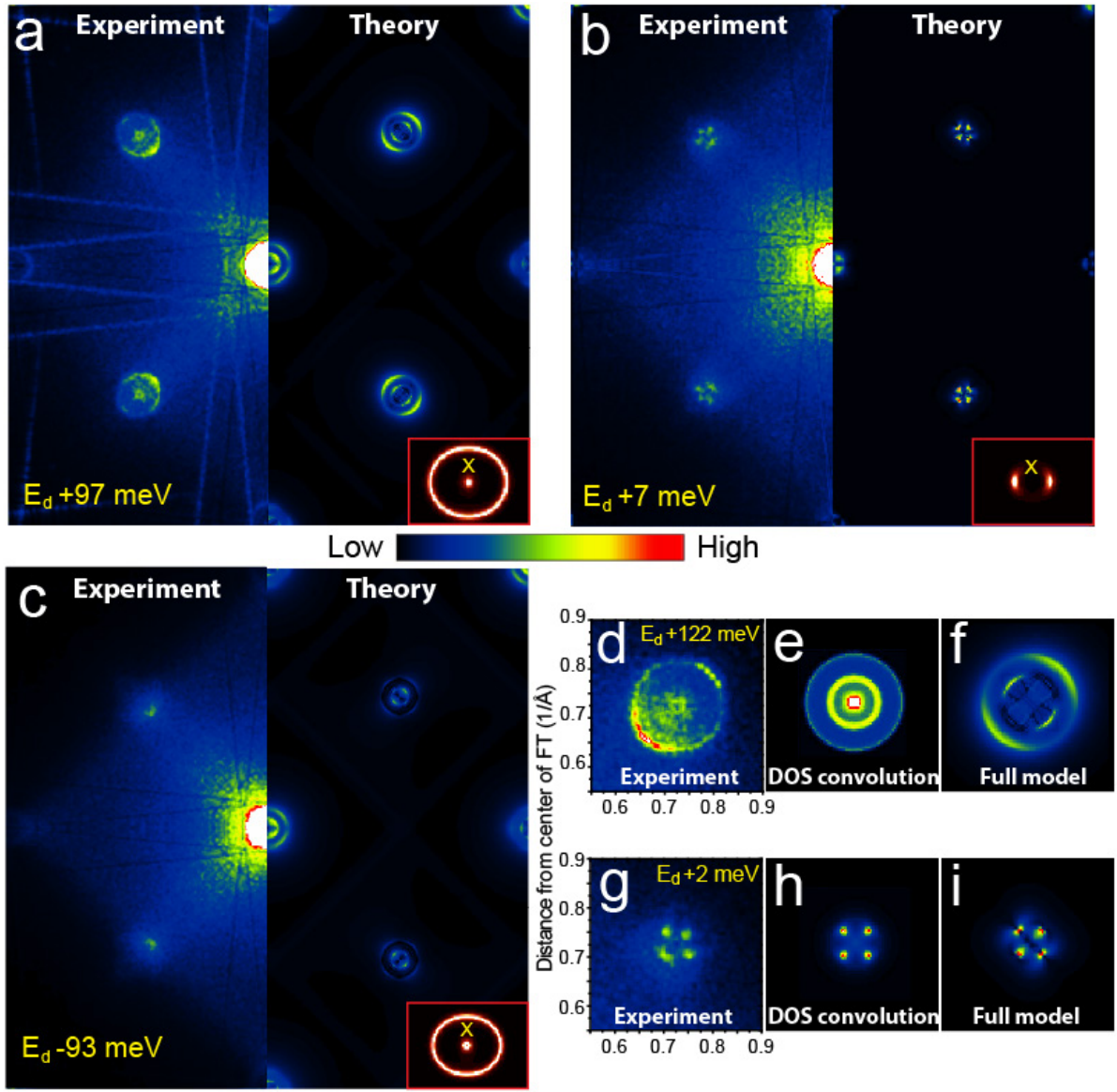

**Figure 4.** Determining the orbital character of the SS bands. Experimentally observed FT-QPI pattern (left half) and calculated pattern (right half) at three representative energies: (a) above the upper Lifshitz transition, (b) close to the Dirac point, and (c) below the lower Lifshitz transition. The pattern in (c) shows suppressed intensity due to the orbital character of the bands as depicted in Fig. 1. Atomic Bragg peaks are located exactly at the corners of the field of view in (a-c). Insets in (a-c) show the CECs used to generate the simulations at each energy around a single X point. Zoom-in of: (d) set of scattering wave vectors $Q_2$ above the upper Lifshitz transition ($E_d$+122 meV), and (g) around the Dirac point energy. (f,i) The simulations at energies corresponding to (d,g), which show an excellent agreement. All experimental images have been 1-pixel boxcar averaged in *q*-space (effectively smoothing over the radius equal to ~0.5% of the atomic Bragg peak wave vector). (e,h) QPI signatures expected from a simple autocorrelation. (e) does not show the radial asymmetry in intensity seen in experiments in (d).